\documentclass[pss]{wiley2sp} 
\usepackage{amsmath}
\usepackage{bm}              
\usepackage{w-greek}         
\usepackage{array}
\newcommand{\pdag}{\phantom{\dag}}

\begin{document}

\title{Quantum criticality in the two-channel pseudogap Anderson model:
A test of the non-crossing approximation}

\titlerunning{Two-channel pseudogap Anderson model}

\author{%
  Farzaneh Zamani,\textsuperscript{\Ast,\textsf{\bfseries 1},\textsf{\bfseries 2}}
  Tathagata Chowdhury,\textsuperscript{\textsf{\bfseries 3}}
  Pedro Ribeiro,\textsuperscript{\textsf{\bfseries 1},\textsf{\bfseries 2}}
  Kevin Ingersent,\textsuperscript{\textsf{\bfseries 3}}
  Stefan Kirchner\textsuperscript{\textsf{\bfseries 1},\textsf{\bfseries 2}}}

\authorrunning{F.\ Zamani et al.}

\mail{e-mail
  \textsf{farzaneh@pks.mpg.de}, Phone:
  +49-351-871-1128}

\institute{%
  \textsuperscript{1}\,Max Planck Institute for the Physics of Complex Systems,
  N\"{o}thnitzer Str. 38, D-01187 Dresden, Germany\\
  \textsuperscript{2}\,Max Planck Institute for Chemical Physics of Solids,
  N\"{o}thnitzer Str. 40, D-01187 Dresden, Germany\\
  \textsuperscript{3}\,Department of Physics, University of Florida,
  Gainesville, Florida 32611-8440, USA
}

\received{XXXX, revised XXXX, accepted XXXX} 
\published{XXXX} 


\abstract{\abstcol{
We investigate the dynamical properties of the two-channel Anderson model
using the noncrossing approximation (NCA) supplemented by numerical
renormalization-group calculations. We provide evidence supporting the
conventional wisdom that the NCA gives reliable results for the standard
two-channel Anderson model of a magnetic impurity in a metal. We extend the
analysis to the pseudogap two-channel model describing a semi-metallic host
with a density of states that vanishes in power-law fashion at the Fermi
energy.}{This model exhibits continuous quantum phase transitions between
weak- and strong-coupling phases.
The NCA is shown to reproduce the correct qualitative features of the
pseudogap model, including the phase diagram, and to yield critical exponents
in excellent agreement with the NRG and exact results. The forms of the
dynamical magnetic susceptibility and impurity Green's function at the
fixed points are suggestive of frequency-over-temperature scaling.}}

\maketitle   

\section{Introduction}

Quantum criticality is currently pursued across many areas of correlated matter,
from insulating or weak magnets to metal-insulator systems to unconventional
superconductors. Interest in continuous phase transitions at temperature $T=0$
is motivated both by the appearance of novel phases near such transitions and
by a richness of quantum critical states that extends beyond the traditional
description in terms of order-parameter fluctuations. A fertile area for
experimental study has been the border of magnetism in rare-earth and actinide
intermetallics, where there is mounting evidence that a crucial issue is the
fate of the Kondo effect on approach to the quantum phase transition
\cite{Si.01,Gegenwart.08}: if Kondo screening becomes critical concomitantly
with the bulk magnetization, then dynamical scaling ensues,
signaling the absence of additional energy scales at criticality
\cite{Schroeder.00,Friedemann.10}.

This paper investigates critical Kondo destruction and dynamical scaling in
the pseudogap two-channel Anderson model. We focus on the treatment of the
problem using the noncrossing approximation (NCA), which involves a
self-consistent evaluation of all irreducible self-energy diagrams without
vertex corrections. Within a pseudoparticle representation of the impurity spin
\cite{Coleman.84}, the NCA threshold exponents of the $T=0$ pseudoparticle
propagators for the one-channel Anderson model \cite{Mueller-Hartmann.84} are
known to differ from the correct exponents inferred from Friedel's sum rule
\cite{Muha.88}. In contrast, the NCA threshold exponents for multichannel
Anderson models agree \cite{Cox.93} with those predicted by boundary conformal
field theory \cite{Affleck.93,Costi.94a}. The NCA is therefore believed to work
well for this model, which has been studied extensively as a relatively simple
route to non-Fermi liquid behavior.
The pseudogap version of the two-channel Kondo model, in which the host density
of states vanishes in power-law fashion at the Fermi energy, was first studied
in Ref.\ \cite{GonzalezBuxton.98}. The model and the counterpart Anderson model
have attracted renewed interest \cite{Schneider.11} due to proposals
\cite{Sengupta.08,Berakdar.10} of their realization in the context of adatoms on
graphene, where the band structure gives rise to two symmetry-inequivalent
Dirac points. Recent scanning tunneling spectroscopy data are in line with
expectations for pseudogap multichannel Kondo physics \cite{Mattos}.

Here we briefly revisit the reliability of the NCA for the metallic two-channel
Anderson model before turning to the pseudogap case, where we apply a scaling
ansatz to obtain $T=0$ threshold exponents, and hence extract critical
exponents describing physical properties. We find good agreement between
these exponents and ones obtained using full numerical solutions of the NCA
equations at $T>0$ and using the numerical renormalization group.


\section{The pseudogap M-channel Anderson model}

The SU$(N)\times$SU$(M)$ Anderson model can be written
\begin{multline}
\label{eq:HANM}
H_{\mathrm{MCA}}
= \sum_{\mathbf{k}}\sum_{\sigma=1}^{N}\sum_{\mu=1}^{M}
   \epsilon_{\mathbf{k}} c_{\mathbf{k}\sigma \mu}^{\dag}
   c_{\mathbf{k}\sigma \mu}^{\pdag} 
+ \epsilon_f \sum_{\sigma} f^{\dag}_{\sigma} f^{\pdag}_{\sigma} \\
   +V\,\sum_{\mathbf{k},\sigma,\mu} \left( c^{\dag}_{\mathbf{k}\sigma\mu}
   f^{\pdag}_{\sigma} b^{\dag}_{\mu} + \mbox{H.c.}\right),
\end{multline}
where $c_{\mathbf{k}\sigma\mu}^{\dag}$ creates a conduction electron of wave
vector $\mathbf{k}$ in channel $\mu$ and spin projection $\sigma$. The
representation $d^{\dag}_{\sigma\mu}=f^{\dag}_{\sigma}b^{\pdag}_{\mu}$
of the impurity electron creation operator in terms of pseudofermion creation
combined with slave-boson annihilation faithfully reproduces the
SU$(N)\times$SU$(M)$ Anderson model provided that the constraint
$Q=\sum_{\sigma} f^{\dag}_{\sigma} f^{\pdag}_{\sigma} + \sum_\mu
b^{\dag}_{\mu} b^{\pdag}_{\mu}=1$ is enforced exactly.

We assume a conduction-electron density of states
\begin{equation}
\label{eq:DOS}
\rho(\omega)=-\frac{1}{\pi}\,\mbox{Im}G_c(\omega)
  =\frac{r+1}{2 D^{r+1}} \, |\omega|^r \, \Theta(D-|\omega|)
\end{equation}
that vanishes at the Fermi energy $\omega=0$ in a manner governed by
exponent $r$, taken to satisfy $0<r<1$; the special case $r=0$
describes a flat (metallic) band. The half-bandwidth $D$ acts as the
basic energy scale in the problem.

$N=2$ and $M=1$, Eq.\ \eqref{eq:HANM} reduces to the standard Anderson impurity
model (with infinite on-site Coulomb repulsion) which in turn contains the
spin-isotropic Kondo model as an effective low-energy model, while for
$N\rightarrow \infty$ various saddle-point approximations can be constructed
\cite{Read.83,Cox.93,Parcollet.98}. The dynamical large-$N$ limit of Ref.\
\cite{Parcollet.98} for the spin-isotropic Kondo model uses a generating
functional equivalent to that for the NCA, making it seem natural to expect
similar results from the two approaches. However, it is important to
note that within NCA, the slave-boson propagator $G_b(\omega)$ is not a
Hubbard-Stratonovich decoupling field and is therefore dynamic even at the bare
level, and that $G_b(\omega)$ couples to the local constraint, which is
enforced exactly. As a consequence, $G_b(\omega)$ and the pseudofermion
propagator $G_f(\omega)$ in the NCA develop threshold behavior reminiscent of
the core hole propagator in the x-ray edge problem, resulting in maximally
particle-hole asymmetric spectral functions:
$\mathrm{Im}\,G_b(\omega\rightarrow 0,T=0)\sim\Theta(\omega)
|\omega|^{-\alpha_b}$ and $\mathrm{Im}\,G_f(\omega\rightarrow 0,T=0)
\sim\Theta(\omega) |\omega|^{-\alpha_f}$. Therefore, $G_b(\tau,T)$ and
$G_f(\tau,T)$ should be rather different from their dynamical large-$N$
counterparts, raising the question of whether dynamical or $\omega/T$
scaling---a property that arises naturally within the dynamical
large-$N$ approach---can carry over to the NCA.

\section{Asymptotically exact zero-temperature solution}

For the case $r=0$ of a constant density of conduction-electron states, an
exact finite-temperature solution can be obtained by transforming the NCA's
integral equations into a set of coupled differential equations
\cite{Mueller-Hartmann.84}. As this procedure relies on specific properties of
the NCA solution for $r=0$, its extension to the pseudogap case is unclear.
We can address the zero-temperature form of the NCA solution by imposing a
scaling ansatz for the pseudoparticle spectral functions. The scaling ansatz
has been successfully used to extract critical properties of impurity models
treated within the dynamical large-$N$ approximation, including the Kondo model
both with a metallic ($r=0$) density of states \cite{Parcollet.98} and with a
pseudogap \cite{Vojta.01}, the Bose-Fermi Kondo model \cite{Zhu.04}, and the
pseudogap Bose-Fermi Kondo model \cite{Zamani.12}. Here we show that a
generalization of this ansatz to the case of extreme particle-hole asymmetry
(generated by the exact enforcement of the constraint) can be used to extract
the critical properties of the fixed points within the NCA.

\begin{figure}[tbp]
\centerline{\includegraphics*[height=\linewidth,angle=90]{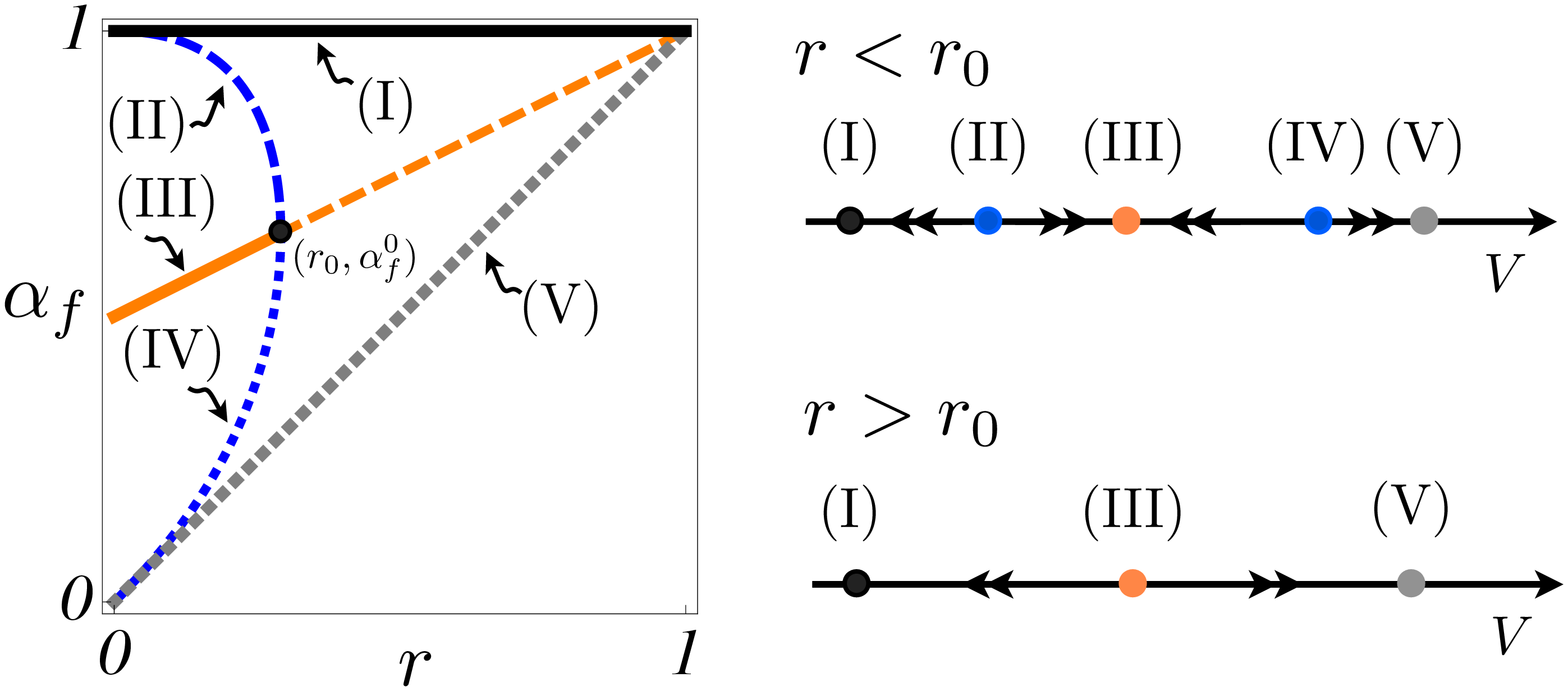}}
\caption{\label{exp}
Left: Variation with pseudogap exponent $r$ of possible solutions $\alpha_f$ of
Eq.\ \eqref{alphaf2} for the two-channel Anderson model. Dashed lines
represent unstable fixed points and dotted lines show values not
observed in numerical solutions of the finite-temperature NCA equations.
Right: Renormalization-group (RG) flows and fixed points as functions of
hybridization $V$ for the ranges $0<r<r_0$ and $r>r_0$. Labels (I)--(V) connect
RG fixed points with NCA solutions in the left panel.}
\end{figure}

The NCA self-energies for the pseudoparticle propagators
at real frequencies are \cite{Kirchner.02,Kirchner.04}
\begin{eqnarray}
\label{NCA1}
\Sigma_{f\sigma}^{\mathrm{ret}}(\omega)
 &=& V^2 \sum_{\mu} \int \! d\epsilon \, f(\epsilon) \,
     A_{c\sigma\mu}(-\epsilon) \, G_b^{\mathrm{ret}}(\epsilon\!+\!\omega), \\
\Sigma_{b\mu}^{\mathrm{ret}}(\omega)
  &=& V^2 \sum_{\sigma} \int \! d\epsilon \, f(\epsilon) \,
      A_{c\sigma\mu}(\epsilon) \, G_f^{\mathrm{ret}}(\epsilon\!+\!\omega),
\label{NCA2}
\end{eqnarray}
where the superscript ``ret'' specifies a retarded function, 
$A_{c\sigma\mu}(\omega)=-\pi^{-1}\mbox{Im}\,
G_{c\sigma\mu}^{\mathrm{ret}}(\omega)=\rho(\omega)$, and $f(\epsilon)$ is
the Fermi-Dirac distribution function. Working at temperature $T=0$, we make
the ansatz \cite{Kuramoto.84,Muha.88}
\begin{equation}
\label{ansatz}
A_p(\omega)  
  =  a_p \: \Theta(\omega) \, \omega^{-\alpha_p}, \qquad p = f, b,
\end{equation}
for the pseudoparticle spectral functions $A_p(\omega)=-\pi^{-1}$ \\
$\mathrm{Im}\,G_p^{\mathrm{ret}}(\omega)$, with $\alpha_p$ being a
threshold exponent and $a_p$ a constant. Substituting these scaling forms into
Eqs.\ \eqref{NCA1} and \eqref{NCA2}, and applying Dyson's equations, one
obtains the self-consistency conditions
\begin{eqnarray}
\label{alphaf1}
\alpha_{f}+\alpha_{b} &=& 1+r, \\
\sin \left(\pi  \alpha _f\right) \sin \left[\pi 
   \left(r-\alpha _f\right)\right]  \times & & \nonumber \\  
   \left\{ \frac{M}{N} \frac{ \sin \left[\pi  \left(r-\alpha
   _f\right)\right]}{\alpha _f}+\frac{\sin \left(\pi 
   \alpha _f\right)}{1+r-\alpha_f}\right\} &=&0.
   \label{alphaf2}
\end{eqnarray}
Equation \eqref{alphaf1} is obtained by matching exponents and Eq.\
\eqref{alphaf2} by matching amplitudes.
Hereafter, we focus on the two-channel case $N=M=2$, for which the possible
solutions of Eqs.\ \eqref{alphaf1} and \eqref{alphaf2} are plotted
schematically on the $r$-$\alpha_f$ plane in Fig.\ \ref{exp}. 
For $r=0$, the solutions are $\alpha_f=0$, $\alpha_f=1$
(local moment), and $\alpha_f=\frac{1}{2}$ (intermediate coupling), in
agreement with Ref.\ \cite{Muha.88}. For $0<r<r_0$, where the condition
$(r_0+1)\pi/2=\cot(r_0 \pi/2)$ yields a numerical value $r_0 \simeq 0.292$,
there are five solutions, of which three correspond to stable
renormalization-group fixed points: local moment (I), two-channel Kondo (III),
and infinite-$U$ resonant level (V). For $r>r_0$, there are just three
solutions, of which only (I) and (V) are stable.

The (physical) impurity Green's function is obtained as
\begin{multline}
G_{d\sigma\mu}^{\mathrm{ret}}(\omega) = \int \! d\epsilon \, e^{-\beta\epsilon}
  \bigl[ G_{f\sigma}^{\mathrm{ret}}(\epsilon\!+\!\omega)\,A_{b\mu}(\epsilon) \\
- G_{b\mu}^{\mathrm{adv}}(\epsilon\!-\!\omega) \, A_{f\sigma}(\epsilon)
  \bigr],
\label{eq:G_d}
\end{multline}
where $\beta=1/T$ and ``adv'' means advanced. The \mbox{$T=0$} scaling form of
this Green's function and of the local spin susceptibility $\chi$ can be deduced from
the pseudoparticle propagators by simple arguments, yielding
$\text{Im}\,G_d^{\mathrm{ret}}(\omega) \propto |\omega|^{1-\alpha_f-\alpha_b}
\equiv|\omega|^{-r}$ and
$\text{Im}\,\chi(\omega)\propto\text{sgn}(\omega)\,|\omega|^{1-2\alpha_f}$.
 
The NRG has been applied previously to the pseudogap two-channel Kondo model
\cite{GonzalezBuxton.98,Schneider.11}. For $r<r_{\mathrm{max}}\simeq 0.23$
(somewhat smaller than the NCA value $r_0\simeq 0.292$), the method finds 
three stable fixed points separated by two unstable critical points,
precisely analogous to the situation shown in Fig.\ \ref{exp}. For
$r_{\mathrm{max}}<r<1$, the NRG yields one critical point between two stable
fixed points, again in one-to-one agreement with the NCA scaling ansatz. Of
these fixed points, only those corresponding to (I)--(III) can be accessed
for level energies $\epsilon_f<0$, the regime considered in the present paper.
(The NCA treatment of cases with $\epsilon_f>0$ will be reported elsewhere
\cite{Zamani.12}.)

We have confirmed that the pseudogap two-channel Anderson model shares the same
fixed points and critical properties as its Kondo counterpart, in agreement
with the conclusions of Ref.\ \cite{Schneider.11}. This justifies our
comparisons below between NCA results for the Anderson model and NRG results
for the Kondo model (the smaller Hilbert space of which allows greater
numerical efficiency).

\section{Finite-temperature NCA solution}

At temperatures $T>0$, the NCA equations are amenable to a numerical solution
on the real frequency axis over a wide parameter range. Details of the numerical
evaluation scheme can be found, \textit{e.g.}, in \cite{Kroha.98}.

\begin{figure}[tbp]
\centerline{\includegraphics*[width=0.8\linewidth]{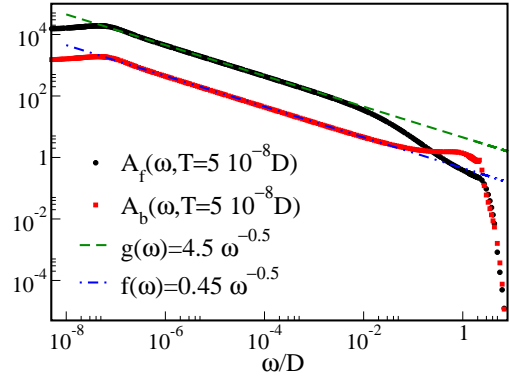}}
\caption{\label{r=0:pspec}
Pseudoparticle spectral functions $A_f(\omega)$ and $A_b(\omega)$ for a
metallic host ($r=0$), calculated at temperature $T=5\times 10^{-8}D$ for
$\epsilon_f/D=-0.6$ and $(V/D)^2=0.5$. The NCA pseudoparticle threshold
exponents $\alpha_f=\alpha_b=0.5$ agree with the exactly known results.}    
\end{figure}

\begin{figure}[tbp]
\centerline{\includegraphics*[width=0.84\linewidth]{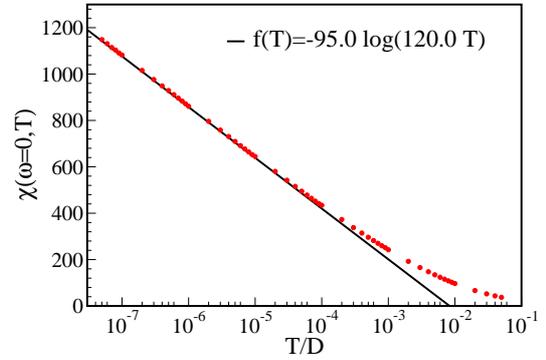}}
\caption{\label{r=0:chi}
Static susceptibility $\chi$ vs temperature $T$ for a metallic host ($r=0$)
and $\epsilon_f/D=-0.6$, $(V/D)^2=0.5$. The NCA captures the logarithmic
divergence of $\chi(T)$.}
\end{figure}

\textbf{Results for r\:=\:0:}
As mentioned above, the NCA predicts the correct threshold exponents for
multichannel Anderson models with a metallic density of states. This is
illustrated in Fig.\ \ref{r=0:pspec}, where the fitted pseudoparticle
exponents agree very well with the NCA scaling ansatz and with the boundary
conformal field theory for the two-channel Kondo model \cite{Affleck.91}.
Figure \ref{r=0:chi} demonstrates that the NCA correctly captures the
logarithmic divergence in temperature $T$ of the local spin susceptibility
$\chi$. The  subleading behavior of the impurity spectral function,
$A_d(\omega)\equiv -\pi^{-1}\mathrm{Im}\,G_d^{\mathrm{ret}}$, is also
reproduced \cite{Cox.93}: $A_d(\omega)-A_d(0)\sim\sqrt{|\omega|}$.
Taken together, these pieces of evidence indicate that the NCA gives
qualitatively correct results for the $r=0$ two-channel Anderson problem.

\begin{table}[t]
\centering
\begin{tabular}{clllll}
& \multicolumn{1}{c}{(I)} & \multicolumn{1}{c}{(II)} & \multicolumn{1}{c}{(III)}
& \multicolumn{1}{c}{(IV)} & \multicolumn{1}{c}{(V)} \\ \hline
$\alpha_f$ & 1.0 & 0.967 & 0.575 & 0.183 & 0.15 \\
$\alpha_b$ & 0.15 & 0.183 & 0.575 & 0.967 & 1.0 \\ \hline
\end{tabular}
\caption{\label{tab:r=0.15}
Pseudoparticle threshold exponents at $T=0$ for the two-channel Anderson model
with $r=0.15$.}
\end{table}

\begin{figure}[tbp]
\centerline{\includegraphics*[width=0.8\linewidth]{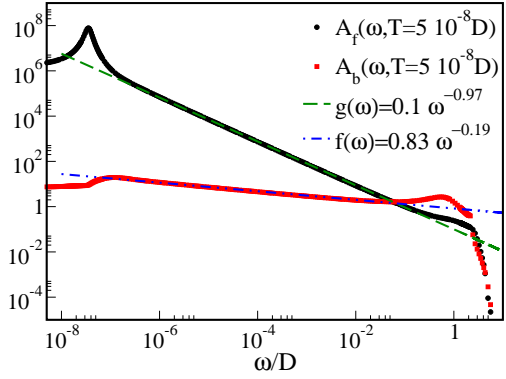}}
\caption{\label{r=0.15b:pspec}
Pseudoparticle spectral functions $A_f(\omega)$ and $A_b(\omega)$ for
$r=0.15$, $T=5\times 10^{-8}D$, $\epsilon_f/D=-0.6$, and $(V/D)^2=0.27$.
The power-law behaviors correspond to solution (II) in Table \ref{tab:r=0.15}
and Fig.\ \ref{exp}.}
\end{figure}

\begin{figure}[tbp]
\centerline{\includegraphics*[width=0.84\linewidth]{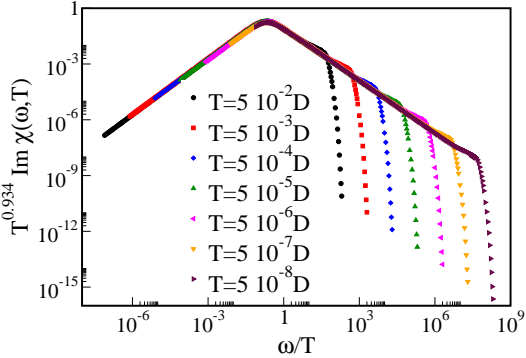}}
\caption{\label{r=0.15b:chi-scaling}
Dynamical scaling of $\chi(\omega,T)$ at the critical fixed point (II) in
Fig.\ \ref{exp}, for $r=0.15$, $\epsilon_f/D=-0.6$, and $(V/D)^2=0.27$.}
\end{figure}

\begin{figure}[tbp]
\centerline{\includegraphics*[width=0.8\linewidth]{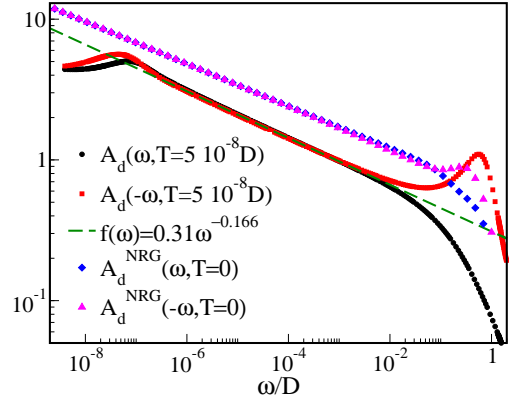}}
\caption{\label{r=0.15b:spec}
Impurity spectral function $A_d(\omega)$ for $r=0.15$ very close to the
critical fixed point (II) in Fig.\ \ref{exp}, calculated using the NCA at
temperature $T=5\times 10^{-8}D$ for $\epsilon_f/D=-0.6$ and $(V/D)^2=0.27$.
Also shown is the spectral function calculated within the NRG for $T=0$,
$\epsilon_f/D=-0.1$ and $(V/D)^2\simeq 0.0486$. The NCA and NRG spectral
functions are described by very similar exponents ($0.166$ and $0.150$,
respectively), and both show that at the critical point, $A_d(\omega)$ is
particle-hole symmetric at low energies.}
\end{figure}

\begin{figure*}
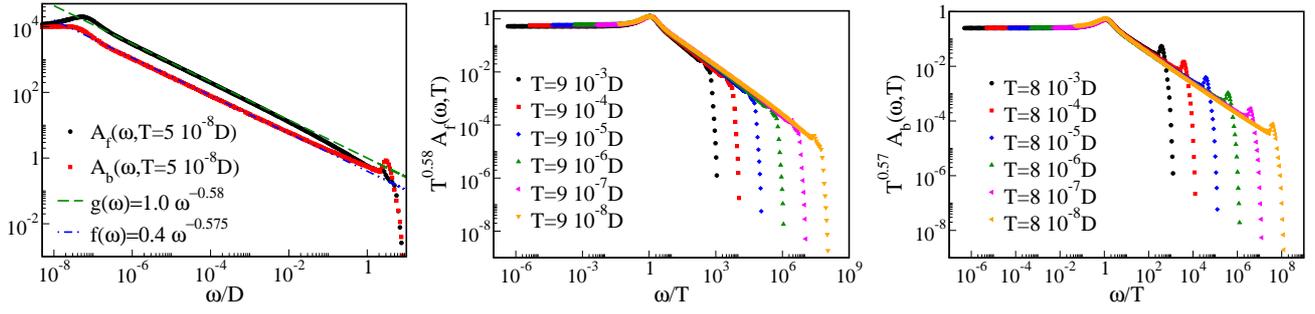

\subfloat{\includegraphics*[width=.31\textwidth]{Figure7a}}\hfill
\subfloat{\includegraphics*[width=.34\textwidth]{Figure7b}}\hfill
\subfloat{\includegraphics*[width=.33\textwidth]{Figure7c}}
\caption{\label{r=0.15c}
Pseudoparticle spectral functions $A_f(\omega,T)$ and $A_b(\omega,T)$
at the fixed point (III) in Fig.\ \ref{exp} for $r=0.15$, $\epsilon_f/D=-0.6$,
and $(V/D)^2=2.0$.
(a) Frequency variation at $T=5\times 10^{-8}$.
(b), (c) Dynamical scaling at different temperatures.}
\end{figure*}

\begin{figure}[tbp]
\centerline{\includegraphics*[width=0.8\linewidth]{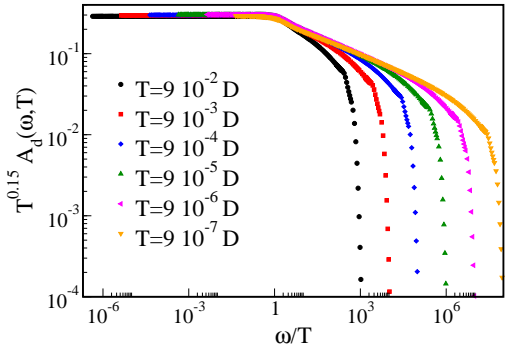}}
\caption{\label{r=0.15c:spec-scaling}
Dynamical scaling of the impurity spectral function $A_d(\omega,T)$ at the
stable fixed point (III) in Fig.\ \ref{exp} for $r=0.15$, $\epsilon_f/D=-0.6$,
and $(V/D)^2=2.0$.}
\end{figure}

\textbf{Results for r\:=\:0.15:}
We now turn to numerical results for the two-channel pseudogap Anderson model,
focusing first on the case $r=0.15$ with $\epsilon_f=-0.6D$ as a representative
example of the behavior in the range of pseudogap exponents $0<r<r_0$.
For this case, Eqs.\ \eqref{alphaf1} and \eqref{alphaf2} predict five solutions,
listed in Table \ref{tab:r=0.15}. Solution (I) corresponds to the local-moment
fixed point where $V$ effectively vanishes.
Obtaining converged numerical solutions near this fixed point is very difficult
as the resulting sharp features cannot be resolved on discrete frequency grids.
Increasing the hybridization $V$ until a solution can be stabilized at the
lowest accessible temperatures yields the critical solution (II) at $V=V_c$
where $(V_c/D)^2\simeq 0.27$. Fig.\ \ref{r=0.15b:pspec} shows that the
pseudoparticle exponents at this critical point agree well with the scaling
ansatz results in Table \ref{tab:r=0.15}. The temperature dependence of the
static susceptibility reflects the frequency behavior. As a result, the
dynamical local spin susceptibility $\chi(\omega,T)$ at the critical point
exhibits the dynamical scaling form (see Fig.\ \ref{r=0.15b:chi-scaling})
\begin{equation}
\label{chi-scaling}
\chi(\omega,T)=T^{-x} \Phi(\omega/T),
\end{equation} 
with $x=0.934$. It is instructive to compare this result against other methods.
The NRG is unable to reliably access the regime $0<|\omega|/T\ll 1$, so it
cannot fully test for dynamical scaling. However, we find for the two-channel
Kondo model with $r=0.15$ that at the critical point, the NRG gives
$\chi(\omega=0,T) \propto T^{-x}$ and
$\mathrm{Im}\,\chi(\omega,T=0)\propto|\omega|^{-y}$ with $x=y=0.930 \pm 0.001$.
These properties are entirely consistent with Eq.\ \eqref{chi-scaling}, and
show that the NCA does an excellent job of calculating the exponent $x$. We
note that this exponent deviates significantly from the leading-order value
$x=1-2r^2=0.955$ coming from an expansion about the local-moment fixed point
\cite{Schneider.11}.

Figure \ref{r=0.15b:spec} shows the impurity spectral function $A_d(\omega)$
very close to the critical fixed point (II) in Table \ref{tab:r=0.15}. The
power-law variation is compatible with the $|\omega|^{-r}$ expected from the
scaling ansatz, which coincides with the exact behavior known to hold at all
intermediate-coupling fixed points \cite{Schneider.11} and with the NRG results
also plotted in Fig.\ \ref{r=0.15b:spec}. Both the NCA and the NRG show the
critical spectral function to be particle-hole symmetric at small $|\omega|$.

\begin{figure}[tbp]
\centerline{\includegraphics*[width=0.8\linewidth]{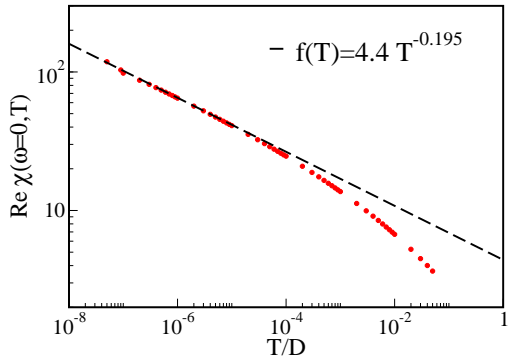}}
\caption{\label{r=0.15c:chi-stat}
Static susceptibility $\chi$ vs temperature $T$ at the stable fixed point (III)
in Fig.\ \ref{exp} for $r=0.15$, $\epsilon_f/D=-0.6$, and $(V/D)^2=2.0$.}
\end{figure}

The stable fixed point (III) in Table \ref{tab:r=0.15} can be accessed by
increasing the hybridization beyond the critical value $V_c$. Figure
\ref{r=0.15c} presents the pseudoparticle spectral functions at various
temperatures for $(V/D)^2=2.0$. The threshold exponents $\alpha_f=0.58$ and
$\alpha_b= 0.575$ extracted from Fig.\ \ref{r=0.15c}(a) are in line with the
prediction $\alpha_f=\alpha_b=(1+r)/2$ of the scaling ansatz. Figures
\ref{r=0.15c}(b) and \ref{r=0.15c}(c) demonstrate that the temperature
dependences of $A_f(\omega,T)$ and $A_b(\omega,T)$ are compatible with the
frequency behavior such that dynamical, or $\omega/T$-scaling ensues. This
carries over to the impurity spectral function, which, as seen in Fig.\
\ref{r=0.15c:spec-scaling}, is compatible with the scaling
\begin{equation}
\label{spec-scaling}
G_d^{\mathrm{ret}}(\omega,T)=T^{-r} \Psi(\omega/T),
\end{equation}
This scaling is consistent with the exact result
\cite{Schneider.11} mentioned above, i.e.,
$A_d(\omega,T=0)\propto|\omega|^{-r}$.

Figure \ref{r=0.15c:chi-stat} plots the static local susceptibility vs
temperature at the stable fixed point (III). A rather narrow window of
asymptotic temperature dependence---presumably a consequence of a strong
subleading contribution to $1/\chi(\omega=0,T)$---gives an exponent $x=0.195$
in reasonable agreement with the scaling ansatz prediction $x=2\alpha_f-1=0.15$.

\begin{figure*}[h!]
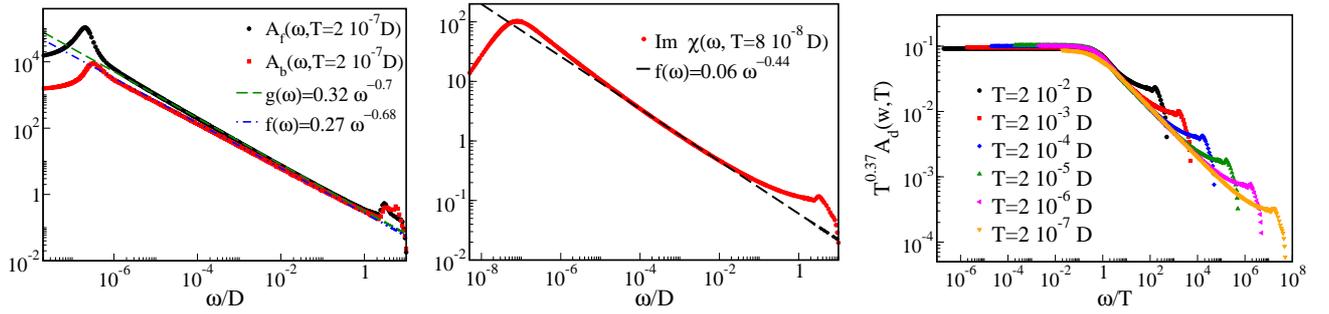
%
\subfloat{\includegraphics*[width=.31\textwidth]{Figure10a}}\hfill
\subfloat{\includegraphics*[width=.315\textwidth]{Figure10b}}\hfill
\subfloat{\includegraphics*[width=.34\textwidth]{Figure10c}}
\caption{\label{r=0.4}
Results for $r=0.4$ in the vicinity of the asymmetric critical point (III) in
Fig.\ \ref{exp}, calculated for $\epsilon_f=-0.55D$ and $(V/D)^2=5.8$.
(a) Pseudoparticle spectral functions $A_f(\omega,T)$ and $A_b(\omega,T)$ at
$T=10^{-8}D$.
(b) Dynamic spin susceptibility $\chi(\omega,T)$ at $T=10^{-8}D$.
(c) Dynamic scaling of the local density of states $A_d(\omega,T)$.}
\end{figure*}

\textbf{Results for r\:=\:0.4:}
We end this section by considering a representative case in the range $r_0<r<1$.
For $r=0.4$, the scaling ansatz predicts an asymmetric critical point described
by $\alpha_f=\alpha_b=(1+r)/2=0.7$, the solution corresponding to (III) in
Fig.\ \ref{exp}. Figures \ref{r=0.4} and \ref{r=0.15c:chi-stat} show results
in the vicinity of this critical point, obtained for $\epsilon_f=-0.55D$ and
$(V/D)^2=5.8$.
The pseudoparticle exponents $\alpha_f=0.7$, $\alpha_b=0.68$ extracted from
Fig.\ \ref{r=0.4}(a) closely follow the scaling ansatz, and the dynamical
scaling of $A_d(\omega,T)$ with an exponent $0.37\simeq r$ [Fig.\
\ref{r=0.4}(b)] agrees with the exact result \cite{Schneider.11}.
Furthermore, the exponent $y=0.44$ of $\chi(\omega,T=0)\propto|\omega|^{-y}$
fitted from the rather narrow frequency window of power-law behavior in Fig.\
\ref{r=0.4}(b) is in reasonable agreement with the value $y=x=2\alpha_f-1=0.4$
predicted by the scaling ansatz under the assumption of dynamical scaling and
with $x=y=0.381 \pm 0.001$ given by the NRG. The impurity spectral function,
shown in Fig.\ \ref{r=0.4}(c), also appears to be consistent with dynamical
scaling.

\section{Conclusion}

We have investigated the reliability of the non-crossing approximation (NCA)
for the two-channel pseudogap Anderson model. This was accomplished by comparing
finite-temperature, finite-frequency solutions of the NCA equations with
asymptotically exact zero-temperature NCA solutions, with numerical
renormalization-group calculations, and with exact results where available.
In contrast to the well-known shortcomings of the NCA for the single-channel
Anderson model with a constant density of states at the Fermi energy, the NCA
captures surprisingly well the asymptotic low-energy properties of the
two-channel model, both for metallic and semi-metallic (pseudogapped) hosts.
In cases of a pseudogap, the results that we have presented for the magnetic
susceptibility and the impurity spectral function are suggestive of 
frequency-over-temperature scaling in the dynamical properties. More complete
testing for dynamical scaling is planned as future work.
Finally, we note that the validation of the NCA treatment of this
problem at equilibrium opens the way for its extension to nonequilibrium
steady-state conditions \cite{Zamani.12}.

\begin{acknowledgement}
S.K.\ acknowledges support under NSF Grant No.\ PHYS-1066293 and the
hospitality of the Aspen Center for Physics. Work at the U.\ of
Florida was supported in part by the NSF MWN program under Grant No.\
DMR-1107814.
\end{acknowledgement}

\providecommand{\WileyBibTextsc}{}
\let\textsc\WileyBibTextsc
\providecommand{\othercit}{}
\providecommand{\jr}[1]{#1}
\providecommand{\etal}{et al.}

\end{document}